\documentclass[sigconf]{acmart}
\usepackage{tikz}
\usepackage{hyperref}
\usepackage{url}
\usepackage{enumitem}
\usepackage{caption}
\AtBeginDocument{%
  }

\setcopyright{acmlicensed}
\copyrightyear{2026}
\acmYear{2026}
\acmDOI{XXXXXXX.XXXXXXX}
\acmConference[32nd ACM SIGKDD Conference (KDD 2026)]{}{August 9--13,
  2026}{Jeju, Korea}
\acmISBN{978-1-4503-XXXX-X/2026/06}

\begin{document}

\title{Isotonic Layer: A Unified Framework for Recommendation Calibration and Debiasing}

\author{Hailing Cheng}
\email{haicheng@linkedin.com}
\orcid{1234-5678-9012}
\affiliation{%
  \institution{Linkedin Inc}
  \city{Mountain View}
  \state{California}
  \country{USA}
}

\author{Yafang Yang}
\email{yafyang@linkedin.com}
\orcid{1234-5678-9012}
\affiliation{%
  \institution{Linkedin Inc}
  \city{Mountain View}
  \state{California}
  \country{USA}
}

\author{Hemeng Tao}
\email{hetao@linkedin.com}
\orcid{1234-5678-9012}
\affiliation{%
  \institution{Linkedin Inc}
  \city{Mountain View}
  \state{California}
  \country{USA}
}

\author{Fengyu Zhang}
\email{fezhang@linkedin.com}
\orcid{1234-5678-9012}
\affiliation{%
  \institution{Linkedin Inc}
  \city{Mountain View}
  \state{California}
  \country{USA}
}

\begin{abstract}
Model calibration and debiasing are fundamental yet operationally expensive challenges in large-scale recommendation systems. Existing approaches treat them as separate problems requiring distinct infrastructure: post-hoc calibration pipelines, propensity estimation workflows, and per-segment model farms. We introduce the \textbf{Isotonic Layer}\footnote{\url{https://github.com/hailingc/Isotonic-Layer}}, a differentiable piecewise-linear module that unifies both problems within a single, lightweight architectural component—requiring no additional data preprocessing, no propensity estimation, and no separate calibration pipelines.

The core insight is elegant: by parameterizing non-negative bucket weights as learnable context embeddings, the model \emph{automatically learns} all calibration and debiasing functions end-to-end from standard training data. Swapping in a different embedding (position, device type, advertiser ID, or any combination) instantly yields calibration tailored to that sub-segment—at arbitrary granularity in any high-dimensional feature space—with no engineering changes beyond a single embedding lookup. The same layer handles post-hoc calibration, position debiasing, and heterogeneous multi-task bias correction within one unified framework. This paper does not claim state-of-the-art over every benchmark; it offers a \emph{principled, practical simplification}: a plug-and-play solution that replaces fragmented, high-maintenance calibration infrastructure with a single end-to-end trainable component. Extensive production A/B tests confirm significant improvements in predictive accuracy, calibration fidelity, and ranking consistency.

\end{abstract}

\begin{CCSXML}
<ccs2026>
   <concept>
       <concept_id>10002951.10003317.10003347</concept_id>
       <concept_desc>Information systems~Recommender systems</concept_desc>
       <concept_significance>500</concept_significance>
   </concept>
</ccs2026>
\end{CCSXML}

\ccsdesc[500]{Information systems~Recommender systems}

\keywords{Multitask Learning, Isotonic Layer, Model Debiasing}

\received{xx mm 2026}
\received[revised]{dd mm 2026}
\received[accepted]{dd mm 2026}

\maketitle

\section{Introduction}
In modern large-scale recommendation systems, the predicted probabilities generated by \textbf{Deep Neural Networks (DNNs)} often diverge from pure latent user preferences. These signals are typically confounded by systemic factors, most notably position bias, presentation bias, and selection bias. While the industry has historically addressed these challenges through post-hoc calibration and debiasing, a fundamental architectural tension persists.

Traditional non-parametric methods, such as \textbf{Isotonic Regression} \cite{leeuw2009}, provide indispensable monotonic guarantees—ensuring that higher predicted relevance corresponds to a higher event probability—but are notoriously difficult to integrate into end-to-end, gradient-based learning. Conversely, standard deep learning layers offer high representational flexibility but lack the global constraints necessary to enforce logically ordered relationships between inputs and outputs.

Previous researchers have explored integrating monotonic properties within deep learning frameworks. For instance, Shen et al. \cite{shen2021framework} introduced \textit{Isotonic Embedding} to represent promotional activities monotonically. However, such approaches primarily utilize the monotonic property of the embeddings without exploring their broader fitting capabilities or utilizing them as universal calibration tools.

The Isotonic Layer was originally conceived by the first authors of this paper and first deployed as a component within \textbf{LiRank}~\cite{lirank2021}, LinkedIn's production ranking system. Due to scope constraints in that system paper, only a global, single-task variant of the layer was described, without formal analysis or generalization to context-conditioned embeddings, multi-task settings, or a unified calibration and debiasing framework. This paper presents the complete formalization of the Isotonic Layer—encompassing its theoretical foundations, monotonicity guarantees, context-conditioned extensions, and dual-tower debiasing formulation—as a unified end-to-end calibration framework that maintains structural monotonicity while leveraging the full representational power of deep ranking models.

\subsection{The Necessity of Monotonicity}
Many recommendation scenarios rely on domain-specific priors regarding the relationship between features and outcomes. For instance, in "feed quality" modeling, an increase in the quality score should, $ceteris paribus$, result in a strictly higher model output. However, in standard deep learning architectures, parameters are updated locally via backpropagation without regard for global order. This lack of architectural coordination often results in "inversion errors"—where Quality Level 5 is scored lower than Quality Level 4 due to data noise—leading to inconsistent rankings and poor generalization in production environments.

\subsection{The Challenge of Task Heterogeneity}
The complexity of debiasing in industrial recommendation engines is compounded by Multi-Task Learning (MTL). Systems must simultaneously optimize for diverse user behaviors with heterogeneous bias profiles; for instance, a "click" is far more sensitive to position and visual prominence than a "purchase."
Current frameworks often fail by treating model outputs as a single, homogeneous score, ignoring these task-specific distortions. Specifically:
\begin{itemize} 
\item \textbf{Rigid Parametric Methods:} Techniques like Platt Scaling lack the capacity to model varying distortion profiles across tasks. 
\item \textbf{Standard Neural Layers:} These lack the global inductive bias necessary to prevent cross-task "inversion" noise. 
\end{itemize}

\subsection{Proposed Solution: The Isotonic Layer}
To address these limitations, we introduce the Isotonic Layer, a differentiable, "plug-and-play" architectural component that bridges the gap between flexible deep learning and rigorous monotonic constraints. Our framework is built upon two core pillars:
\begin{itemize}
\item \textbf{Differentiable Piecewise Fitting:}
We partition the input feature space into discrete segments, each governed by a local fitting weight $w_i$. By enforcing a non-negativity constraint ($w_i \ge 0$) through an activation function (e.g., ReLU or Softplus), we instantiate a global inductive bias that guarantees the output is monotonically non-decreasing.
\item \textbf{Debiasing as Calibration:}
We reframe debiasing as a monotonic transformation problem. By parameterizing segment weights as learnable, context-aware embeddings, the model "learns the distortion" for specific features like rank or task type. This allows the Isotonic Layer to act as a universal debiasing agent, reshaping distorted scores back into calibrated utility scores while preserving the underlying model's ranking integrity.
\end{itemize}

\subsection{Design Philosophy}
The central value of this work is \emph{unification through simplicity}. Calibration and debiasing are not treated as separate problems requiring separate infrastructure—they are both instances of the same learned monotonic transformation, automatically derived from training data without any manual data engineering. Specifically:
\begin{itemize}
\item \textbf{No data preprocessing required.} Unlike IPS-based methods that need propensity estimation or randomized logging, the Isotonic Layer learns bias patterns directly from observational data.
\item \textbf{Arbitrary sub-segment calibration via a single embedding change.} Calibrating for a new context (e.g., a specific advertiser, device, or position) requires only adding an embedding lookup—no new model, no new pipeline.
\item \textbf{Fully automatic, end-to-end.} All calibration and debiasing functions are learned by the model itself during standard training. No post-hoc fitting, no manual curve construction.
\item \textbf{Plug-and-play for any multi-task setup.} Task-specific embeddings handle heterogeneous bias profiles across arbitrary objectives within one unified architecture.
\end{itemize}

\subsection{Contributions}
The key innovations of this work are summarized as follows:
\begin{enumerate}
\item \textbf{Unified Calibration \& Debiasing Framework:}
A single differentiable layer replaces the need for separate calibration pipelines, propensity models, and per-task sub-models, unifying both problems under one end-to-end trainable component.
\item \textbf{Context-Conditioned Isotonic Embeddings:}
Bucket weights conditioned on arbitrary context features (e.g., advertiser ID, device type, display position) enable fine-grained per-segment calibration at any granularity in high-dimensional spaces—impossible with classical methods.
\item \textbf{Handling Task Heterogeneity:}
Task-specific isotonic embeddings learn heterogeneous bias profiles for distinct objectives (CTR vs.\ CVR) within a unified MTL pipeline, dynamically accounting for the unique distortion intensities of each engagement type.
\item \textbf{Dual-Task Formulation:}
We decouple recommendation into latent relevance estimation and bias-aware calibration, using the Isotonic Layer as a differentiable functional bridge, with bias fully neutralizable at inference without retraining.
\end{enumerate}
Mathematically, for an input $x$ and fixed knots $k_0, \dots, k_n$, the layer computes:$$ y = w_0(k_1 - k_0) + w_1(k_2 - k_1) + \dots + w_i(x - k_i) $$This piecewise linear approach provides sufficient capacity to prevent underfitting while maintaining the structural properties required for robust, unbiased recommendations in heterogeneous environments.

\section{Related Work}

\textbf{Classical calibration.}
\textbf{Platt Scaling}~\cite{platt1999probabilistic} applies a parametric logistic transformation to predicted scores but is too restrictive for non-linear distortions in modern recommenders. \textbf{Isotonic Regression}~\cite{zadrozny2002transforming} offers a non-parametric monotonic alternative but relies on non-differentiable projection (PAVA), resists end-to-end training, and produces staircase mappings that degrade ranking resolution.

\textbf{Differentiable monotonic calibration.}
The Isotonic Layer was originally conceived by its first authors (Cheng and Yang) and deployed within \textbf{LiRank}~\cite{lirank2021} (patent US20250278623A1) as a single-task, global instantiation; this paper provides its complete formal treatment and generalization. \textbf{DIPN}~\cite{dipn2020} similarly uses isotonic embeddings for monotonic incentive–response modeling. Both apply a single global function and target single-task settings. Unconstrained monotonic networks~\cite{umnn2023calibration} improve expressiveness but remain limited to score-level calibration without addressing task heterogeneity or context-specific bias.

\textbf{Causal and counterfactual debiasing.}
\textbf{Position-as-Feature (PAF)} and \textbf{Inverse Propensity Scoring (IPS)}~\cite{joachims2017unbiased} reweight samples to simulate randomized exposure, but suffer from inference-time distribution shift and high propensity estimation variance, respectively. Recent approaches model bias through disentangled representations~\cite{dbvae2023}, adversarial learning~\cite{apwcf2025}, or meta-optimization (\textbf{AutoDebias}~\cite{Chen_2021}), and address multifactorial biases~\cite{huang2024multifactorial,mansoury2026unfairness,huang2024dclmdb,ipm2025invariantdebias,carnovalini2025popularity,klimashevskaia2024popularitysurvey}. These methods operate on data distributions or latent representations without imposing structural constraints on score transformations.

\textbf{Multi-task debiasing.}
Prior work mitigates negative transfer via task decoupling or causal regularization~\cite{orca2022}, but most methods assume a shared calibration mechanism across tasks despite large differences in exposure sensitivity across objectives such as clicks versus conversions.

\textbf{Our approach} treats calibration and debiasing as the \emph{same} differentiable monotonic transformation, unified in a single plug-and-play layer requiring no preprocessing or propensity estimation. Context- and task-conditioned isotonic embeddings enable fine-grained per-objective calibration within a unified multi-task framework—addressing bias heterogeneity that global monotonic or causal reweighting methods do not handle.

\section{Isotonic Layer Architecture Overview}
\label{sec:isotonic_layer}

\subsection{Two Applications of the Same Layer}
\label{sec:unified_recipe}

The Isotonic Layer is a single architectural component that serves two distinct applications in recommendation systems: \textbf{calibration} (Section~\ref{sec:calibration}) and \textbf{debiasing} (Section~\ref{sec:debiasing}). In both cases, the mechanism is the same—an embedding lookup on a context feature parameterizes the isotonic bucket weight vector, and the model learns the corresponding monotonic transformation end-to-end. The important difference lies in \emph{how} the layer is deployed across training and serving:

\begin{itemize}
  \item \textbf{Calibration:} the Isotonic Layer takes calibration context features (e.g., platform type, user segment) as input and is active during \emph{both} offline training and online serving. The same context-conditioned calibration function learned during training is applied identically at inference, continuously correcting systematic score distortions without any additional pipeline.
  \item \textbf{Debiasing:} the Isotonic Layer takes bias-inducing context features (e.g., feed position, device type) as input and forms a dedicated \emph{Isotonic Calibration Head} trained alongside a standard \emph{Inference Head}. It is active only during offline training, where it provides a corrective learning signal that shapes the shared representation. At online serving time, the Isotonic Head and its debiasing features are \emph{removed entirely}; only the Inference Head is deployed, producing rankings driven by pure latent relevance free of positional or device bias.
\end{itemize}

The train/serve asymmetry reflects the fundamental difference: calibration distortions persist at serving time and must be corrected there, whereas debiasing signals (position, device) are logging-policy artifacts that must not influence online predictions. In both cases the underlying building block is identical, requiring no additional pipeline, preprocessing, or propensity estimation.

\begin{figure}[t]
  \centering
  \includegraphics[width=0.8\columnwidth]{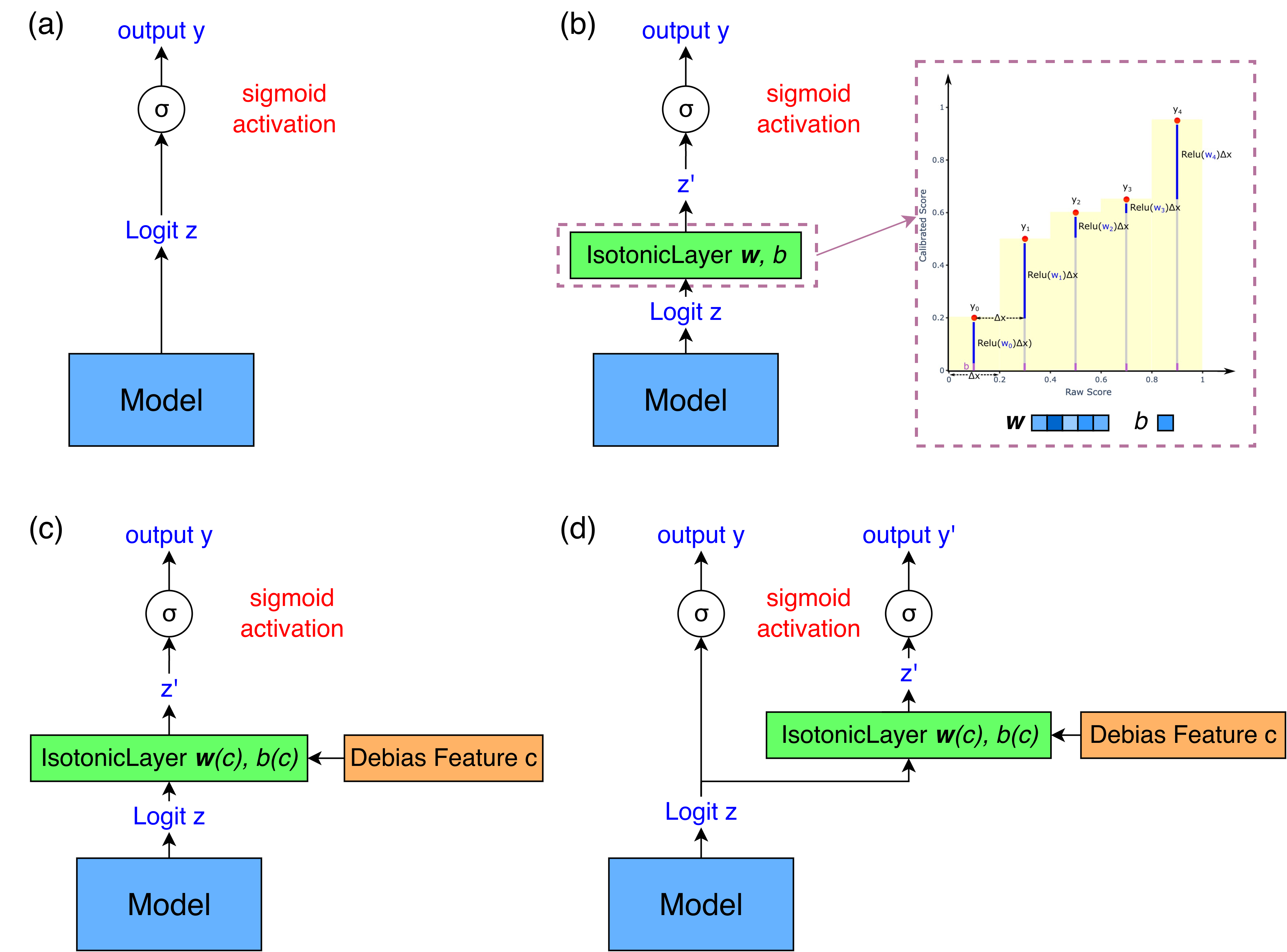}
  \caption{
  Example use cases of the Isotonic Layer in ranking and calibration.
  (a) Standard prediction model without monotonic calibration.
  (b) Global isotonic calibration applied to model outputs with learnable
  bucket weights $\mathbf{w}$ and bias $\mathbf{b}$.
  (c) Context-conditioned isotonic calibration via learned embeddings, where a context feature $c$ is mapped to a
  context-conditioned isotonic embedding $(\mathbf{w}_c, b_c)$.
  (d) Dual-tower architecture with joint optimization of relevance and
context-aware isotonic calibration for debiasing.
  }
  \label{fig:isotonic_usecase}
\end{figure}

We propose an \emph{Isotonic Layer} (Figure~\ref{fig:isotonic_usecase}(b)): a differentiable, bucket-based neural module that enforces monotonically non-decreasing outputs via cumulative bucket activations and ReLU-constrained non-negative weights, fully compatible with end-to-end gradient-based optimization.

\subsection{Mathematical Formulation}

Let $x \in \mathbb{R}$ denote a scalar input.
The Isotonic Layer maps $x$ to an output $y \in (0,1)$ such that for any
$x_1 \le x_2$, the corresponding outputs satisfy $y(x_1) \le y(x_2)$.

\paragraph{Input Clipping.}
The input is first clipped to a predefined range $[L, U]$:
\begin{equation}
\tilde{x} = \mathrm{clip}(x, L, U),
\end{equation}
where $L$ and $U$ denote the lower and upper bounds, respectively.

In our implementation, we set the default bounds to
$L=-17$, $U=8$, with a bucket width $\Delta_b = 0.05$.
These values are not theoretically unique, but are chosen as
practical defaults based on empirical observations in large-scale
click-through rate (CTR) and engagement prediction tasks. \footnote{
For the sigmoid function $\sigma(z) = (1 + e^{-z})^{-1}$, saturation occurs when the
derivative $\sigma'(z) = \sigma(z)(1 - \sigma(z))$ becomes negligibly small.
At $z = -17$, $\sigma(-17) \approx 4.1 \times 10^{-8}$ and
$\sigma'(-17) \approx 4.1 \times 10^{-8}$, while at $z = 8$,
$\sigma(8) \approx 0.999665$ and $\sigma'(8) \approx 3.35 \times 10^{-4}$.
Beyond this interval, both the output probability and its gradient change by less than
$10^{-4}$ per unit change in $z$, implying that further variation in the logit has
negligible effect on the output or optimization dynamics. Clipping to this range therefore
introduces minimal approximation error while improving numerical stability and calibration
robustness.}

\paragraph{Bucketization.}
The interval $[L, U]$ is discretized into $N$ buckets of width $\Delta_b$:
\begin{equation}
N = \left\lceil \frac{U - L}{\Delta_b} \right\rceil + 1 .
\end{equation}
The clipped input $\tilde{x}$ is mapped to a bucket index
$i \in \{0, 1, \dots, N-1\}$ via
\begin{equation}
i = \left\lfloor \frac{\tilde{x} - L + \Delta_b}{\Delta_b} \right\rfloor ,
\end{equation}
with $i$ clamped to the valid range $[0, N-1]$.

\paragraph{Activation Vector Construction.}
For each input $\tilde{x}$, an activation vector $\mathbf{a}(\tilde{x}) \in \mathbb{R}^N$
is constructed to represent the accumulated contribution of buckets:
\begin{equation}
a_j(\tilde{x}) =
\begin{cases}
\Delta_b, & j < i, \\
(\tilde{x} - L + \Delta_b) - i \cdot \Delta_b, & j = i, \\
0, & j > i.
\end{cases}
\end{equation}
The fractional term captures the partial activation within the current bucket.

\paragraph{Weighted Aggregation.}
Each isotonic unit maintains a learnable weight vector
$\mathbf{w} \in \mathbb{R}^N$ and bias $b \in \mathbb{R}$.
To enforce monotonicity, weights are parameterized as
\begin{equation}
\mathbf{w}^+ = \mathrm{ReLU}(\mathbf{w}),
\end{equation}
ensuring $\mathbf{w}^+ \ge 0$ element-wise.
The pre-activation output is computed as
\begin{equation}
z(x) = \sum_{j=0}^{N-1} a_j(\tilde{x}) \, w^+_j + r + b,
\end{equation}
where $r = L - \Delta_b$ is a constant residue offset for boundary alignment.

\paragraph{Output Activation.}
The final output is obtained by applying a sigmoid transformation:
\begin{equation}
y(x) = \sigma(z(x)) = \frac{1}{1 + e^{-z(x)}}.
\end{equation}

\paragraph{Loss Function and Training.}
The Isotonic Layer is trained end-to-end using standard task-specific loss
functions applied to the sigmoid output $y(x)$.
Given a labeled dataset $\{(x_i, t_i)\}_{i=1}^M$, where $t_i \in \{0,1\}$ denotes
the target label (e.g., click or engagement), we minimize a conventional
probabilistic loss such as Binary Cross Entropy (BCE):
\begin{equation}
\mathcal{L}_{\mathrm{BCE}} =
- \frac{1}{M} \sum_{i=1}^M
\left[
t_i \log y(x_i) + (1 - t_i) \log (1 - y(x_i))
\right].
\end{equation}
The layer can be trained jointly with the upstream model or on a separate holdout set; the latter decouples calibration from relevance learning, improving robustness under distribution shift.

\subsection{Monotonicity Guarantee}

For any $x_1 \le x_2$, the activation vectors satisfy $\mathbf{a}(x_1) \le \mathbf{a}(x_2)$ element-wise by construction. Since $w^+_j = \mathrm{ReLU}(w_j) \ge 0$, the dot product preserves this ordering: $z(x_1) \le z(x_2)$, and the strictly monotonic sigmoid yields $y(x_1) \le y(x_2)$. Monotonicity is therefore guaranteed by construction—it is precisely the \textbf{ReLU} on bucket weights that enforces non-negativity and thereby isotonicity.

\paragraph{Generalization: Unconstrained Piecewise Linear Fitting.}
The ReLU constraint is not always necessary. In applications where a monotonicity guarantee is not required—for example, fitting a general score transformation or modeling non-monotonic feature interactions—the ReLU can simply be removed, allowing bucket weights $w_j$ to take arbitrary real values. The layer then becomes a \textbf{universal piecewise linear function approximator}, capable of fitting any continuous shape over the input domain. This makes the same architecture directly applicable as a general-purpose learned transformation beyond the calibration and debiasing settings described in this paper.

\subsection{Expressiveness and Empirical Illustration}

\begin{figure}[t]
  \centering
  \includegraphics[width=0.8\columnwidth]{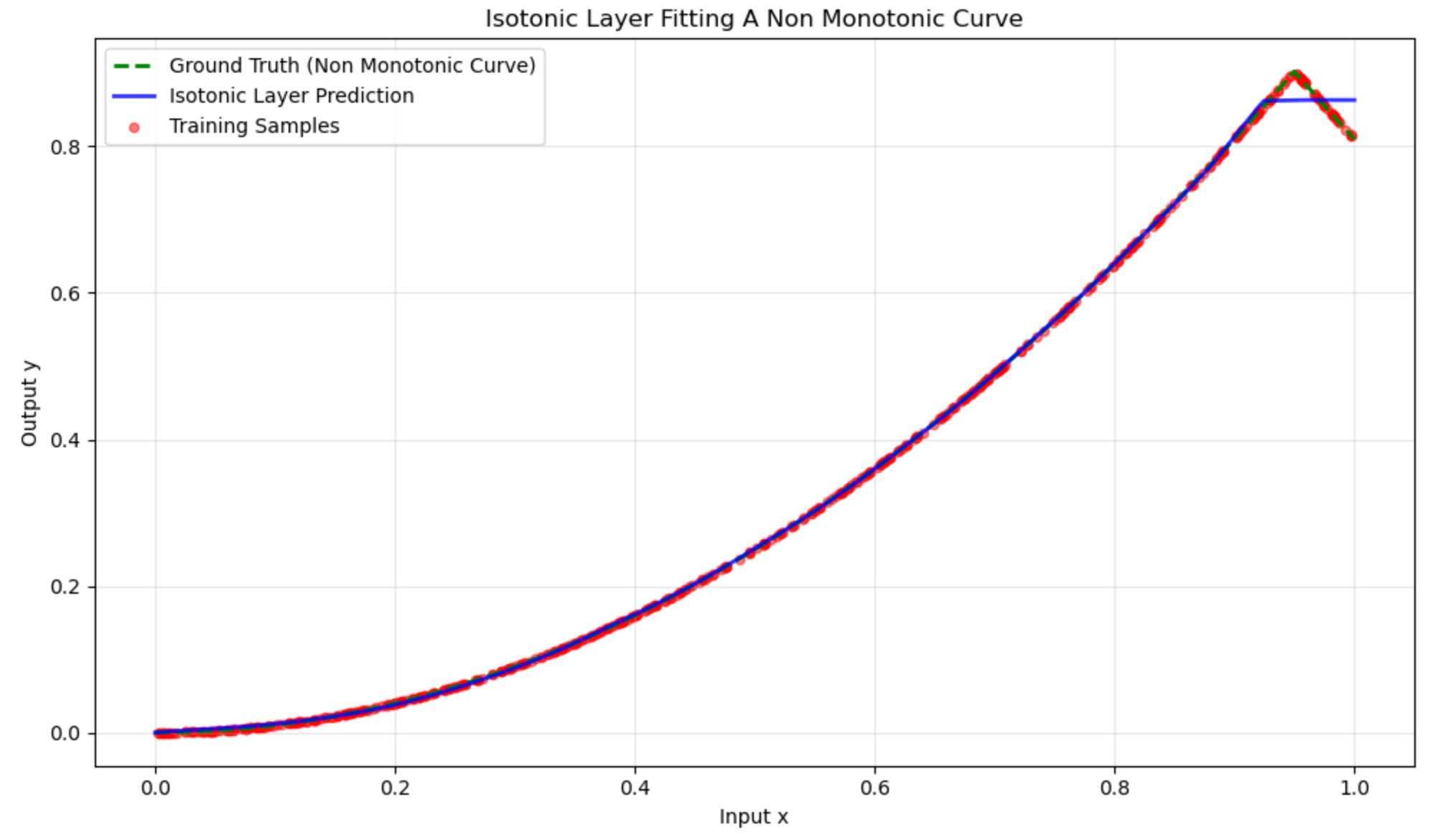}
  \caption{Robustness to non-monotonic noise. The Isotonic Layer enforces global monotonic
  structure while smoothing local distortions in the target signal.}
  \Description{Demo of monotonic task2}
  \label{fig:isotonic}
\end{figure}

Figure~\ref{fig:isotonic} shows the layer achieving near-perfect approximation of $y = x^2$ (inputs in logit space, optimized with BCE), confirming strong expressive power under monotonic constraints. Figure~\ref{fig:isotonic2} evaluates robustness to non-monotonic distortions via a piecewise target ($y = x^2$ for $x \le 0.95$; $y = (1.9-x)^2$ for $x > 0.95$): the layer suppresses local inversions while recovering the optimal monotonic fit, demonstrating structural regularization under noisy inputs.

\subsection{Global Structure vs.\ Local Parametrization}

Unlike MLPs whose parameters act locally, the Isotonic Layer imposes a global structural constraint: the output at $x$ depends on the cumulative prefix of all preceding bucket weights, inducing spatial coupling across segments. This acts as an implicit regularizer—preventing local overfitting or monotonicity violations even under data sparsity—while retaining high expressive capacity through fine-grained bucketization.

\subsection{Context-Conditioned Isotonic Embeddings}

The bucket weights serve as a \textbf{learnable embedding} parameterizing a monotonic transformation (Figure~\ref{fig:isotonic_usecase}(c)). For a context variable $c$, an embedding lookup or lightweight projection yields $\mathbf{w}^+(c) = \mathrm{ReLU}(\mathbf{E}(c)) \in \mathbb{R}^N$, giving a distinct isotonic calibration curve per context while preserving monotonicity:
\begin{equation}
y = \mathrm{IsotonicLayer}(x \,;\, \mathbf{w}^+(c),\, b(c)).
\end{equation}
For example, associating each display position $p$ with embedding $\mathbf{E}_p$ captures position-specific score distortions as $\hat{y} = f_{\mathrm{iso}}(r;\,\mathbf{E}_p)$, where $r$ is the raw relevance score. (See Appendix~\ref{appendix:implementation} for implementation details.)

\subsection{Dual-Tower Debiasing Architecture}

As illustrated in Figure~\ref{fig:isotonic_usecase}(d), the model decomposes into a \textbf{Relevance Tower} estimating latent utility $r$ and an \textbf{Isotonic Calibration Layer} mapping $r$ to the observed interaction space via context $c$:
\begin{equation}
\hat{y} = f_{\mathrm{iso}}(r \,;\, c).
\end{equation}
At inference, bypassing the calibration layer yields rankings driven purely by latent utility ($\hat{r} \approx r$), enabling counterfactual evaluation and fair ranking without retraining.

\section{Isotonic Layer as a Calibration Framework}
\label{sec:calibration}
This section demonstrates the first instantiation of the unified recipe (Section~\ref{sec:unified_recipe}): using the Isotonic Layer for \emph{calibration} by supplying a platform-type context feature. Here, the Isotonic Layer with its context embedding is active during both offline training and online serving—the same calibration function that is learned during training is applied identically at inference, continuously correcting systematic score distortions without any additional pipeline.

In many real-world applications, models suffer from significant overfitting and severe score shifts when training data is sparse. We encounter these challenges specifically in downstream modeling tasks, where labels represent deep conversions—such as offsite purchases in advertising or long-term retention—that are statistically sparse yet high in business value. In this section, we evaluate the Isotonic Layer as a standalone calibration framework for such tasks.

Our specific use case involves estimating the conditional probability of downstream session, defined as the probability that a member returns to the platform via a notification after an initial feed interaction (e.g., like, comment, or share):
$$P(\text{downstream session} \mid \text{user action})$$
This prediction is critical for optimizing notification triggering and ranking policies in large-scale recommender systems. By treating this task as a calibration problem, we leverage the monotonic constraints of the Isotonic Layer to stabilize predictions against the volatility inherent in sparse label distributions, thereby mitigating the score shifts common in traditional deep learning approaches.

\subsection{Motivation}

The downstream engagement model is trained on relatively sparse and
highly conditional data, as downstream sessions occur significantly less
frequently than initial engagement events.
As a result, the model is prone to overfitting and exhibits high variance
in predicted probabilities, particularly in the tail regions of the
score distribution.

In production, this instability manifests as undesirable score swings
across model iterations and sensitivity to data shifts.

\subsection{Calibration via Isotonic Layer}

To address these challenges, we utilize the proposed Isotonic Layer as a robust post-hoc calibration module. We implement an \textbf{alternating training procedure} that explicitly decouples representation learning from probability calibration. 

The training process consists of two distinct, serialized phases:
\begin{itemize}
    \item \textbf{Representation Learning:} The primary downstream model is trained on engagement data to generate raw probability estimates $\hat{y}_{raw}$.
    \item \textbf{Calibration Optimization:} The base model parameters are frozen, and the Isotonic Layer is optimized over these static outputs using randomized session data to learn the mapping $f: \hat{y}_{raw} \to \hat{y}_{calib}$.
\end{itemize}

By enforcing a monotonic constraint on the raw model outputs, the Isotonic Layer ensures that final probabilities remain stable and consistent. This architectural decoupling effectively mitigates the volatility and score shifts prevalent in data-sparse regimes, providing a more reliable signal for downstream decision-making. In this setting, the Isotonic Layer operates as a global calibration module without any context feature input—a single shared isotonic function is learned across all samples. To account for varying behavior across different contexts, we employ independent Isotonic Layers for distinct tasks and user segments, such as mobile versus desktop platforms.

\subsection{Experiment Results}
We evaluate the \textbf{Isotonic Layer} within a production downstream model using product data. The baseline is a \textbf{multi-task learning (MTL)} model that optimizes several binary objectives via shared MLP layers and task-specific towers.
\paragraph{Offline Ranking Performance} We define downstream engagement metrics based on the causal impact of a user's session actions on subsequent platform activity. Specifically, a $Like$, $Share$, or $Comment$ action is assigned a positive label $y=1$ if it triggers at least one downstream session via notifications; otherwise, $y=0$. As detailed in Table~\ref{tbl:ds_auc}, the treatment model yields significant improvements in  \textbf{Evaluation AUC}, with gains of  \textbf{+1.5\%} for Downstream Share and  \textbf{+1.9\%} for Downstream Comment. These results suggest that the monotonic constraints enforced by the Isotonic Layer provide a robust inductive bias, effectively mitigating overfitting in data-sparse regimes where traditional unconstrained models struggle to generalize.
\begin{table}[ht]
  \caption{Relative Performance Improvement of Evaluation AUC}
  \centering
  \begin{tabular}{|lccc|} \hline
    Model & Share Eval AUC & Like Eval AUC &  Comment Eval AUC\\ \hline
    Baseline & - & - & -\\ \hline
    Treatment & \textbf{+1.5\%} & +0.0\% & \textbf{+1.9\%}\\ \hline
\end{tabular}
\label{tbl:ds_auc}
\end{table}

\begin{figure}[ht]
    \centering
    \includegraphics[width=1.0\linewidth]{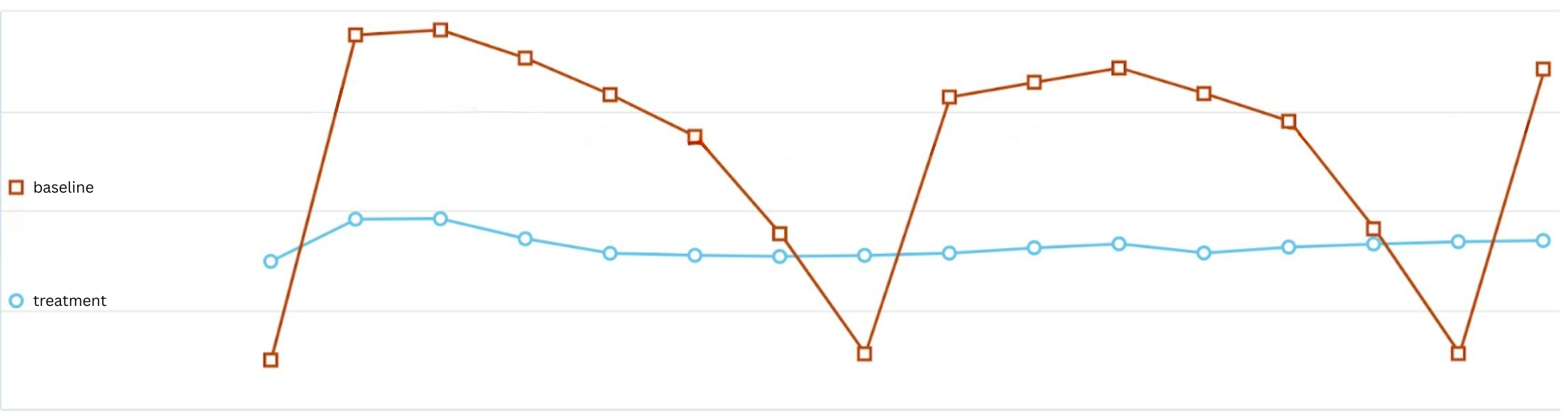}
    \caption{Comparison of Online Prediction Scores. It monitors model average output scores in a daily basis, where X axis represents date and Y axis represents the probabilities.}
    \label{fig:ds_score}
\end{figure}

\paragraph{Prediction Stability and Calibration} Online monitoring highlights the Isotonic Layer’s capability in stabilizing model predictions. As illustrated in \textbf{Figure~\ref{fig:ds_score}}, the treatment model exhibits significantly lower variance in daily average scores compared to the baseline. Furthermore, the post-calibration scores are notably lower, aligning more closely with the empirical distribution of the training labels and correcting the inherent overestimation typical of uncalibrated MTL models.

\begin{table}[ht]
\centering
\small
\caption{Online A/B Test Results: Relative Lift over Baseline. Metric names are anonymized per LinkedIn's external publication policy; categories are described in the text.}
\label{tab:online_ab_results}
\begin{tabular}{lc}
\toprule
\textbf{Metric Category} & \textbf{Relative Lift ($\Delta$ \%)} \\ \midrule
User Engagement (daily active interactions) & \textbf{+0.17\%} \\
Content Quality (knowledge-seeking sessions) & \textbf{+0.20\%} \\
Ecosystem Health (top-line business metric) & \textbf{+0.36\%} \\ \bottomrule
\end{tabular}
\label{tb:ds_metrics}
\end{table}

\paragraph{Online A/B Testing} Live traffic experiments yield statistically significant lifts across all three metric categories as reflected in \textbf{Table~\ref{tb:ds_metrics}}: user engagement (+0.17\%), content quality (+0.20\%), and ecosystem health (+0.36\%). All reported gains are statistically significant at $p < 0.05$ under a two-sided t-test applied by LinkedIn's production A/B testing framework. Metric names are anonymized per company policy; category descriptions are provided in the table footnote. \footnote{Specific metric names are withheld per LinkedIn's competitive disclosure policy. The three categories correspond to: (1) daily active user interactions with feed content; (2) sessions involving professional knowledge-seeking or advice-seeking behavior; (3) a top-line ecosystem health metric sensitive to business strategy.}

\section{Isotonic Layer Debiasing Architecture and Experiments}
\label{sec:debiasing}
This section demonstrates the second instantiation of the unified recipe (Section~\ref{sec:unified_recipe}): using the Isotonic Layer for \emph{debiasing} by conditioning on bias-inducing context features—feed position and device type—during offline training only. At online serving time, the Isotonic Calibration Head and its debiasing features are removed; only the Inference Head is deployed, yielding rankings based on pure latent relevance.

We implement the Isotonic Layer within a multi-task dual-tower framework. This design allows for high-fidelity debiasing during training while maintaining low-latency inference. The Isotonic Calibration Head is conditioned on two debiasing context features: \textbf{feed position} (the rank slot at which an item was displayed) and \textbf{device type} (e.g., iPhone, Android, or PC), whose embeddings are jointly used to parameterize the context-conditioned isotonic bucket weights.

\subsection{Dual-Tower Architecture}
\begin{figure}[ht]
  \centering
  \includegraphics[width=1.0\columnwidth]{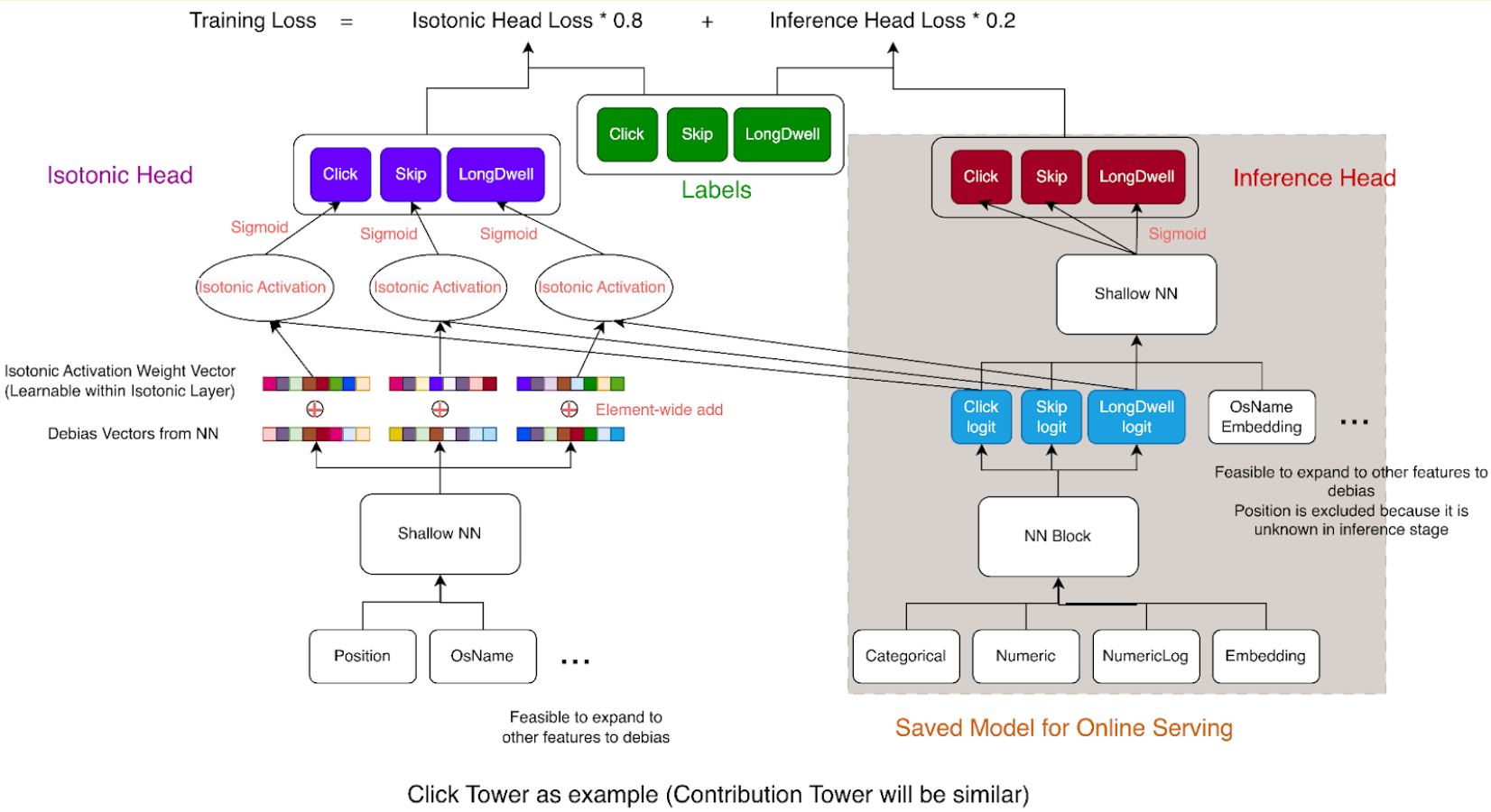}
  \caption{Dual-head architecture for isotonic position debiasing. Each task group consists of a position-neutral inference head and a context-conditioned isotonic head. The isotonic head utilizes position and platform features during training to decouple exposure bias from intrinsic relevance. Loss weights for each head are optimized as hyperparameters to balance ranking performance and model stability.}
  \Description{User A and User B input sequence}
  \label{fig:model_arch}
\end{figure}

The model utilizes a shared latent representation layer followed by task-specific prediction heads. As shown in \textbf{Figure \ref{fig:model_arch}}, for each engagement signal $s \in \{\text{Click, Skip, LongDwell}\}$, we deploy two parallel towers:
\begin{itemize}
\item \textbf{Inference Head:} Predicts a position-neutral relevance score $\hat{y}_{\text{inf}}^{(s)}$. This head is used exclusively during online serving.
\item \textbf{Isotonic Calibration Head:} Learns a bias-aware probability $\hat{y}_{\text{iso}}^{(s)}$ by applying the Isotonic Layer to the shared representation, conditioned on display position and platform features.
\end{itemize}

\subsection{Joint Optimization}The system is optimized using a weighted Binary Cross-Entropy (BCE) loss:$$\mathcal{L} = \sum_{s} \left( \alpha_s \text{BCE}(y^{(s)}, \hat{y}_{\text{inf}}^{(s)}) + \beta_s \text{BCE}(y^{(s)}, \hat{y}_{\text{iso}}^{(s)}) \right)$$
Through empirical tuning, we found that a relative weight of $0.7 \leq \beta_s \leq 0.8$ for the isotonic head yields the best debiasing results while ensuring the inference head remains well-calibrated for downstream value functions.

\subsection{Baselines}

Beyond the production baseline, we include \textbf{Position Dropout} as a competitive debiasing baseline. Position Dropout is a variant of the Position-as-Feature (PAF) paradigm: display position is included as an explicit input feature during training, and a stochastic dropout mask is applied to the position embedding at rate $p_{\text{drop}}$, replacing the position with a sentinel value of $-1$, to prevent the model from overfitting entirely to positional signals. At inference time, the position feature is fully masked ($-1$), so that the model must generalize to position-agnostic predictions.

We performed a systematic hyperparameter search over $p_{\text{drop}} \in \{0.05, 0.10, 0.15, 0.20, 0.30, 0.50\}$. A dropout rate of $p_{\text{drop}} = 0.15$ consistently produced the strongest offline AUC improvement and was selected for all reported comparisons. Despite this tuning, Position Dropout did not yield statistically significant gains in online A/B testing under any evaluated configuration, motivating the need for the structured debiasing approach proposed in this paper.

Prior to this work, the LinkedIn feed ranking team also evaluated \textbf{Platt Scaling} and a suite of \textbf{12 per-task Isotonic Regression models} (one per engagement objective). Both approaches yielded negative to flat top-line metrics in online experiments and introduced substantial architectural complexity—requiring separate training pipelines, model versioning, and serving infrastructure for each task—and were consequently not adopted in production. Due to paper length constraints, detailed metrics for these experiments are omitted, but their failure to deliver production gains further motivates the unified end-to-end framework proposed in this work.

\subsection{Offline and Online Performance}
We evaluate the inference head's ability to isolate $P(\text{relevance})$ by benchmarking against the production baseline and the Position Dropout baseline. \textbf{Table~\ref{tab:system_results}} summarizes the offline evaluation gains.

\begin{table}[h]
\centering
\caption{Offline performance comparison across $Like$ and $Click$ tasks. Position Dropout uses $p_{\text{drop}}=0.15$ (best tuned) with position$=-1$ at inference. Isotonic and Inference heads are components of the proposed dual-tower architecture.}
\label{tab:offline_results}
\begin{tabular}{llcc}
\hline
\textbf{Task} & \textbf{$\Delta$ AUC (\%)} & \textbf{$\Delta$ NE (\%)} \\ \hline
Baseline & - & - \\
Position Dropout Like Task & $-0.34$\% & $+0.21$\% \\
Position Dropout Click Task & $+0.16$\% & $+0.08$\% \\
Isotonic Head's Like Task  & \textbf{+0.81}\% & \textbf{-0.51}\% \\
Isotonic Head's Click Task  & \textbf{+1.00}\% & \textbf{-1.31}\% \\
Inference Head's Like Task  & \textbf{+0.10}\% & $+0.21$\% \\
Inference Head's Click Task  & \textbf{+0.01}\% & $+0.06$\% \\ \hline
\end{tabular}
\label{tab:system_results}
\end{table}

\subsection{Online A/B Testing Results}

The online results represent the impact of the positional debiasing model on live traffic metrics.
\begin{table}[h]
\centering
\caption{Online A/B Testing Results (Production Environment). Position Dropout ($p_{\text{drop}}=0.15$) showed no statistically significant lift on any top-tier metric across all evaluated configurations. The Isotonic Model results correspond to the inference head used during online serving. All reported lifts are statistically significant at $p < 0.05$ under a two-sided t-test.}
\label{tab:online_results}
\begin{tabular}{lc}
\hline
\textbf{Model / Metric} & \textbf{$\Delta$ Relative to Control} \\ \hline
Position Dropout (all top-tier metrics) & $\approx 0$ (not significant) \\
\hline
\multicolumn{2}{l}{\textit{Isotonic Model (Inference Head):}} \\
Subscription Weekly Active User  & \textbf{+0.63}\% \\
Daily Unique Professional Interactions & \textbf{+0.14}\% \\
Job Session & \textbf{+0.14}\% \\
Total Macrosessions & \textbf{+0.06}\% \\ \hline
\end{tabular}
\end{table}

\subsection{Analysis and Key Takeaways}

Experimental results demonstrate that the \textbf{Isotonic Calibration Layer} effectively decouples relevance from position bias, yielding consistent ranking improvements across diverse engagement signals.

\begin{itemize}
\item \textbf{Effective Bias Decomposition:} The model isolates $P(\text{relevance})$ from $P(\text{event} \mid \text{relevance, position})$, resulting in \textbf{AUC} gains. While we observe a slight regression in \textbf{Normalized Entropy (NE)}, this is a mathematically expected and deliberate consequence of the debiasing design. To see why, consider that a conventional model trained without an explicit position feature implicitly learns $\hat{P}(\text{action} \mid \text{features, position})$ which effectively absorbing positional effects into the score. The inference head, by design, targets only $P(\text{relevance} \mid \text{features})$, discarding the positional component. As a result, its scores will systematically \emph{under-predict} for items observed at top positions (where $P(\text{action} \mid \text{position})$ is inflated by exposure) and \emph{over-predict} for items at lower positions. This divergence from the observed action distribution directly degrades NE — not because the model is less accurate about true relevance, but because NE is computed against position-confounded labels. The NE regression is therefore a \emph{sign of successful debiasing}, not a failure of the model.
\item \textbf{Position Dropout Fails to Generalize Online:} Despite systematic offline hyperparameter tuning—with $p_{\text{drop}} = 0.15$ producing the best offline results—Position Dropout did not yield statistically significant gains in any online A/B experiment. We attribute this to the inherent \emph{train-inference distribution mismatch} of the PAF paradigm: the model is trained with stochastically masked position features but tested with a fixed sentinel value (position $= -1$), creating a covariate shift the model cannot fully compensate for. A low dropout rate still allows position overfitting during training, while a high dropout rate degrades position-aware calibration. The Isotonic Layer avoids this dilemma by making the bias-relevance decoupling explicit and structurally enforced.
\item \textbf{Heterogeneity and Directionality of Bias:} Bias is non-uniform and task-specific; for instance, \textit{Click} signals exhibit severe positive bias at top ranks, while \textit{Skip} signals exhibit a reciprocal trend. We also observe a systematic divergence in the \textbf{Observed-to-Expected (O/E) ratio} ($O/E > 1$ at top positions; $O/E < 1$ at lower ranks), proving that a static propensity score cannot resolve the non-linearities captured by our isotonic approach. Detailed O/E ratios by feed position for both the Baseline and Positional Debias Model are reported in Appendix~\ref{appendix:oe_ratio}.
\item \textbf{Loss Weighting Optimization:} The relative loss weights ($\alpha_s, \beta_s$) are critical for system stability. Empirically, a calibration weight $\beta_s \in [0.7, 0.8]$ yields optimal online performance. Retaining a non-zero inference weight $\alpha_s$ is essential to ensure predicted scores remain well-calibrated, preventing volatility in the downstream value function and maintaining ranking consistency.
\end{itemize}

\subsection{Deployment Considerations}

\paragraph{Training vs.\ serving.}
During training, both heads are jointly optimized: the Isotonic Calibration Head consumes position and device features, and its gradients flow into the shared representation, forcing it to disentangle relevance from positional exposure. At serving time, the Isotonic Head and its debiasing features are removed entirely; only the Inference Head is deployed, producing position-neutral relevance scores with no serving-time distribution shift.

\paragraph{CPU overhead and distillation.}
Retaining the Isotonic Head at inference incurs $\approx$\textbf{5\% CPU overhead} ($N \approx 125$ buckets, 12 task heads). Where latency is binding, the head's learned mapping can be distilled into a shallow MLP with negligible accuracy loss ($<0.1\%$ AUC degradation), preserving the full debiasing benefit in the Inference Head.

\section{Future Work}

We identify several promising directions to extend the Isotonic Layer for more complex recommendation environments:

\begin{itemize}
\item \textbf{Multivariate and Partial Isotonicity:} Extending constraints to multiple dimensions (e.g., quality, recency) to ensure joint monotonicity across complex feature sets.
\item \textbf{Contextual Calibration:} Parameterizing the layer with high-dimensional embeddings to learn specialized calibration curves for millions of sub-segments (e.g., specific advertisers) within a unified model.
\item \textbf{Position Action Prediction (PAP):} Implementing a secondary task to predict position distributions $P(\text{pos} = i \mid \text{relevance})$, enabling soft-attention pooling over position embeddings to bridge the exposure gap between training and inference.

\item \textbf{Advanced Off-Policy Evaluation (OPE):} Leveraging the interaction between latent utility and positional bias to generate high-fidelity propensity scores for robust off-policy evaluation.
\end{itemize}

\section{Conclusion}
Industrial recommendation systems today maintain sprawling calibration infrastructure: separate propensity estimation pipelines, per-segment isotonic regression model farms, and task-specific post-hoc correction stages—each with its own engineering overhead, data dependency, and failure mode. The \textbf{Isotonic Layer} replaces all of this with a single differentiable architectural component.

The core contribution is not a new state-of-the-art benchmark result, but a \emph{principled unification}: model calibration, position debiasing, and multi-task bias correction are all instances of the same learned monotonic transformation, automatically derived end-to-end from standard training data with no additional preprocessing. Calibrating any new sub-segment—a specific advertiser, device, position slot, or any combination—requires only pointing an embedding lookup at the new context feature. The model learns the rest. This plug-and-play design works across any multi-task setup without architectural redesign and without any propensity estimation or data curation overhead.

Our extensive experimental results on real-world datasets, supplemented by successful large-scale production A/B tests, demonstrate that the Isotonic Layer significantly enhances calibration fidelity and ranking consistency without compromising model expressivity. The practical impact of this simplification is significant: the approach has enabled teams to retire dozens of separately maintained calibration models while simultaneously improving top-line business metrics.

A further practical advantage, particularly salient at industrial scale, is the compatibility of the Isotonic Layer with \textbf{incremental training}. Because calibration and debiasing are learned end-to-end from standard training data, the model continuously adapts its calibration and bias correction as new data arrives—with no need to re-engineer training pipelines, massage incoming data into a special format, or recompute Inverse Propensity Score (IPS) statistics for each new data batch. IPS-based methods, by contrast, require a fresh propensity estimation cycle whenever the data distribution shifts, introducing a costly offline preprocessing dependency that breaks the incremental training loop. The Isotonic Layer has no such dependency: it treats a new batch of training data identically to any other, making it a natural fit for the continuous, large-scale model update pipelines that underpin modern production recommender systems.

As industrial recommendation systems grow in complexity, the Isotonic Layer offers a robust, model-agnostic foundation—one layer, all tasks, any context, end-to-end.

\bibliographystyle{ACM-Reference-Format}
\bibliography{base}

\appendix
\section{Observed-to-Expected Ratio Analysis}
\label{appendix:oe_ratio}

The Observed-to-Expected (O/E) ratio measures calibration quality at the position level: an O/E of 1.0 indicates perfect calibration, O/E $> 1$ indicates the model \emph{under-predicts} (observed engagement exceeds model expectation), and O/E $< 1$ indicates the model \emph{over-predicts} (model expectation exceeds observed engagement). Systematic deviations from 1.0 across positions are a direct signature of uncorrected position bias.

Table~\ref{tab:oe_ratio} reports O/E ratios for the Baseline Model and the Positional Debias Model across representative feed positions (0–3 and 6) from online A/B testing. Positions 4 and 5 are reserved for advertisements in LinkedIn's feed and are therefore excluded from this analysis: because ad slots follow a different impression and engagement distribution governed by auction dynamics rather than organic ranking, their O/E ratios do not reflect organic recommendation quality and would confound the position-bias measurement.

\begin{table}[h]
\centering
\caption{Online A/B Test: Observed-to-Expected (O/E) Ratio by Feed Position. Values closer to 1.0 indicate better calibration. O/E $>$ 1 = under-prediction; O/E $<$ 1 = over-prediction.}
\label{tab:oe_ratio}
\begin{tabular}{lcc}
\hline
\textbf{Feed Position} & \textbf{Baseline Model} & \textbf{Positional Debias Model} \\ \hline
0 & 1.070 & 1.060 \\
1 & 0.983 & 0.996 \\
2 & 1.020 & 1.010 \\
3 & 0.942 & 0.941 \\
6 & 0.848 & 0.952 \\ \hline
\end{tabular}
\end{table}

The results reveal two patterns. First, the baseline model exhibits a characteristic position-bias signature: O/E $> 1$ at top positions (0, 2) and O/E $< 1$ at lower positions (1, 3, 6), reflecting the tendency of an undebiased model to absorb positional exposure effects into its score predictions — over-predicting engagement at lower-ranked positions that are systematically under-exposed, and under-predicting at top positions where exposure inflates observed rates beyond true relevance.

Second, the Positional Debias Model produces O/E ratios consistently closer to 1.0 across all evaluated positions. The most substantial improvement occurs at position 6, where the baseline O/E of 0.848 (over-prediction by 15.2\%) improves to 0.952 (over-prediction reduced to 4.8\%), confirming that the Isotonic Calibration Head effectively learns and corrects lower-position bias during training. The improvements at positions 1 and 2 are also notable, while position 3 remains similar across both models. These results corroborate the qualitative O/E divergence discussed in Section~\ref{sec:isotonic_layer} and provide quantitative evidence that the Isotonic Layer's position-conditioned embeddings learn meaningful, position-specific calibration corrections.

It is important to note, however, that O/E ratio is not a direct or sufficient measurement of debiasing quality in production recommendation systems. O/E quantifies how well a model's predicted probabilities match observed engagement rates at each position slot, but it does not directly capture whether the model is producing better relevance-ordered rankings. In practice, a model could achieve O/E $\approx 1$ at every position simply by predicting the marginal engagement rate for each slot — with no relevance discrimination at all. The ultimate arbiter of debiasing quality remains top-line business metrics: whether users engage more meaningfully, discover more relevant content, and have better platform experiences overall. What effective position debiasing does contribute is surfacing more genuinely relevant posts into top positions — posts that would have been suppressed in a position-biased model due to lower historical exposure — thereby improving the quality of what users see first and driving the top-line engagement gains reported in Table~\ref{tab:online_results}.

\section{Learned Context-Conditioned Calibration Curves}
\label{appendix:calibration_curves}

The curves presented in this section are extracted directly from the \textbf{Isotonic Calibration Head} of the trained dual-tower debiasing model described in Section~\ref{sec:debiasing}. After training converges on LinkedIn's production feed ranking data, we freeze the model and enumerate the learned calibration functions by sweeping over context feature combinations—feed position, device platform (iOS, Android, Web), and engagement task (Click, Skip)—and passing each combination as the context embedding input to the Isotonic Calibration Head. No post-processing or manual fitting is applied; the curves are the direct output of the learned isotonic bucket weights for each context.

Each curve plots the raw inference score $\sigma(\text{logit})$ on the $x$-axis against the calibrated engagement probability $\sigma(\text{calibration\_logit})$ on the $y$-axis. Taken together, the curves expose the true complexity of the joint calibration problem over the space $[\text{task} \times \text{position} \times \text{platform} \times \text{logit}]$ in real production data. The patterns are highly non-linear and non-additive across dimensions: the interaction between task, position, and platform cannot be decomposed into independent per-dimension corrections, and the functional form varies qualitatively across context combinations rather than merely shifting in scale.

This complexity directly motivates the end-to-end isotonic approach. Classical alternatives cannot address it: \textbf{Platt Scaling} has only two free parameters and fits a single global linear transformation across all contexts; \textbf{per-task isotonic regression} handles one dimension at a time and is blind to the cross-dimensional interactions visible here; \textbf{IPS} assigns a scalar propensity weight per sample but does not learn a calibration surface over the joint context space. All three are forced to average over the structurally distinct regimes these curves reveal, producing a single approximation that fits none of them well. The Isotonic Layer resolves this by learning the full $[\text{task} \times \text{position} \times \text{platform} \times \text{logit}]$ calibration structure end-to-end from standard training data, via a single context-conditioned embedding lookup with no additional pipeline, no manual curve construction, and no propensity estimation.

\begin{figure}[h]
  \centering
  \includegraphics[width=1.0\columnwidth]{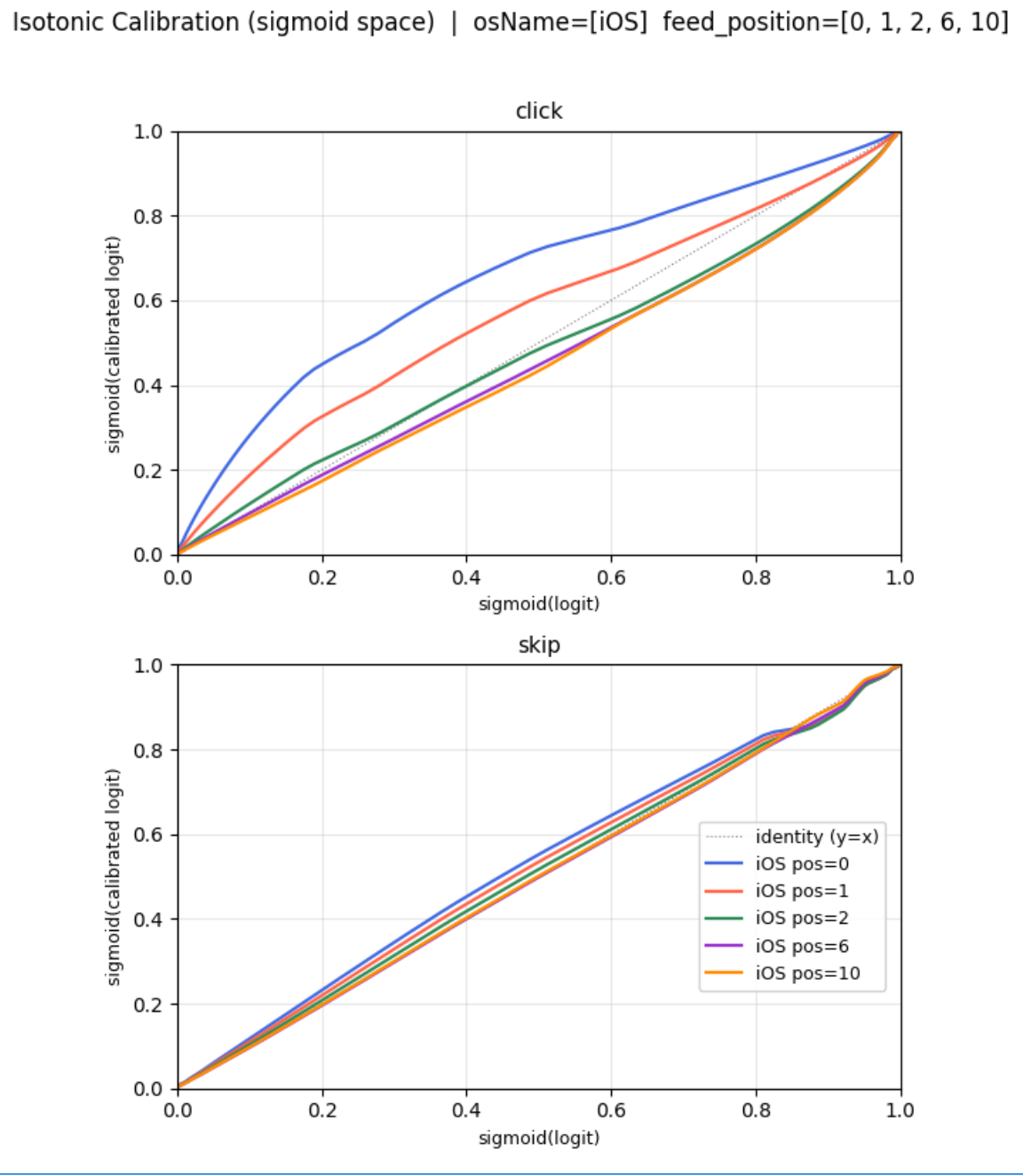}
  \caption{Calibration curves extracted from the production Isotonic Calibration Head (Section~\ref{sec:debiasing}), shown by task and feed position on iOS.
  The $x$-axis is the raw inference score $\sigma(\text{logit})$; the $y$-axis is the
  calibrated probability $\sigma(\text{calibration\_logit})$ output by the Isotonic Head.
  Curves are shown for \textit{Click} and \textit{Skip} tasks at feed positions
  $\{0, 1, 2, 6, 10\}$.
  The substantial divergence between tasks and across positions confirms that position
  bias is both task-specific and position-specific: \textit{Click} exhibits a strong
  upward shift at top positions (high exposure inflation), while \textit{Skip} displays
  a qualitatively different and less position-sensitive pattern.
  No global propensity score, per-task isotonic regression, or Platt Scaling curve
  can capture this heterogeneity; the Isotonic Layer learns all curves end-to-end
  directly from production training data.}
  \label{fig:task_curves}
\end{figure}

\begin{figure}[h]
  \centering
  \includegraphics[width=1.0\columnwidth]{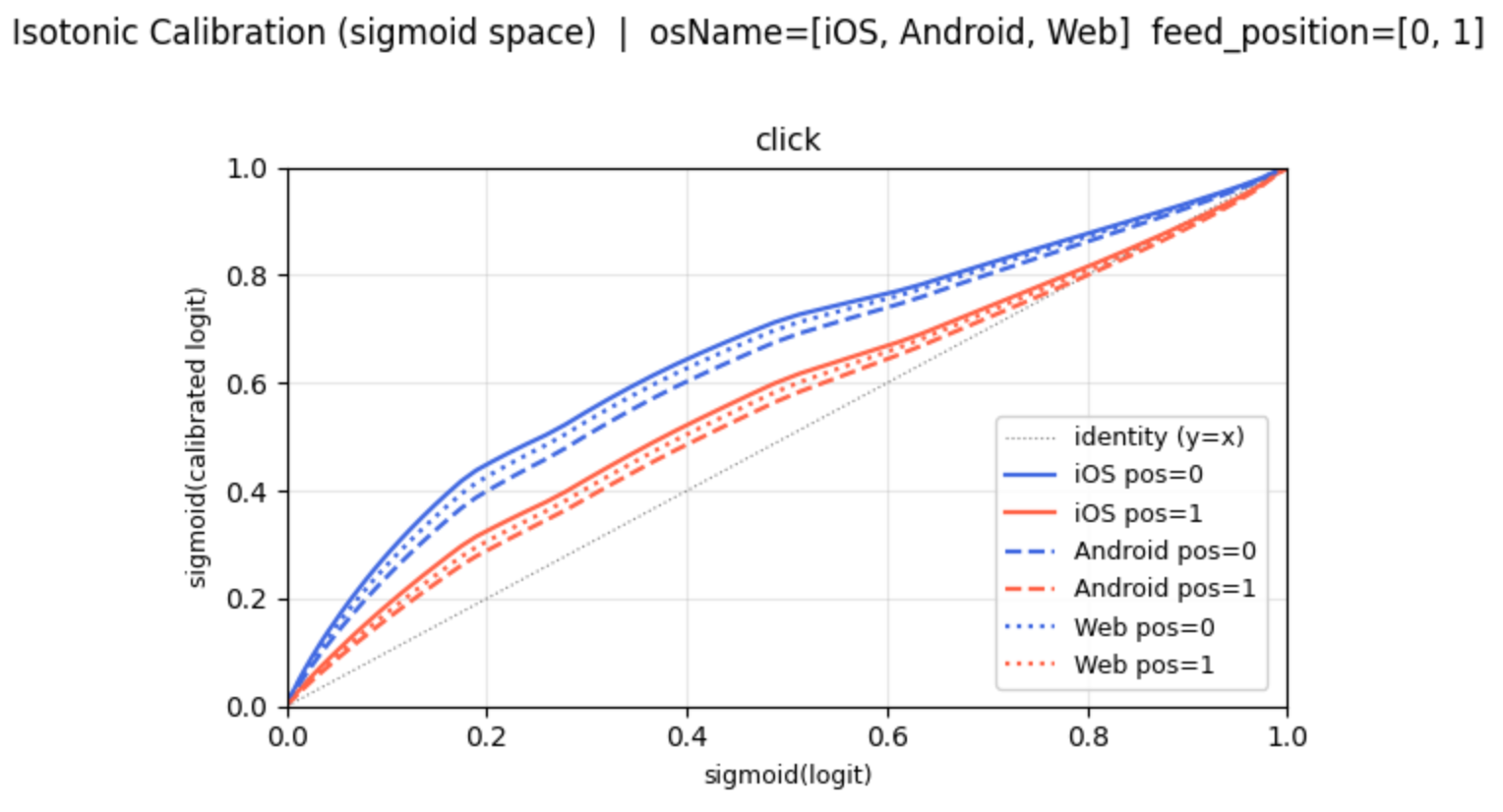}
  \caption{Calibration curves extracted from the production Isotonic Calibration Head (Section~\ref{sec:debiasing}), shown by platform for the \textit{Click} task
  at feed positions 0 and 1.
  Platforms iOS, Android, and Web each produce distinct calibration curves, reflecting
  differences in display layout, screen real estate, and user interaction patterns that
  cause the same raw relevance score to map to different observed engagement rates.
  Position 0 curves sit above position 1 curves within each platform, quantifying the
  position-exposure effect; the gap magnitude varies across platforms, confirming that
  positional and platform biases interact non-additively—a joint structure that
  per-platform or per-position models cannot represent.
  The Isotonic Layer captures the full $[\text{task} \times \text{position} \times \text{platform} \times \text{logit}]$ structure via a single context-conditioned
  embedding lookup, with no platform-specific model farms or separate pipelines.}
  \label{fig:platform_curves}
\end{figure}

\section{Implementation Details}
\label{appendix:implementation}

The Isotonic Layer is designed to be a "plug-and-play" component. By enforcing non-negativity on weights, we ensure a monotonic relationship through a piecewise linear fitting. Below, we provide a PyTorch implementation that leverages optimized tensor operations for efficiency.

\begin{small}
\begin{verbatim}
import torch
import torch.nn as nn
import torch.nn.functional as F

class IsotonicLayer(nn.Module):
    """
    Isotonic Layer implementation in PyTorch.
    Ensures monotonicity via piecewise linear
    fitting with non-negative weights.
    """
    def __init__(
        self,
        units=12, # unit is the task number
        lb=-17.0, # lower bound for logits
        ub=8.0, # upper bound for logits
        step=0.2,
        w_init_factor=0.1
    ):
        super(IsotonicLayer, self).__init__()
        self.units = units
        self.lb = lb
        self.ub = ub
        self.step = step
        self.num_buckets = int((ub - lb) / step) + 1
        self.residue = lb - step

        # Learnable parameters
        self.w = nn.Parameter(
            torch.ones(units, self.num_buckets)
            * w_init_factor
        )
        self.b = nn.Parameter(torch.zeros(units))

    def forward(self, x, calibration_embedding=None):
        # x shape: [batch_size, units] or [batch_size, 1]
        if x.dim() == 1:
            x = x.unsqueeze(1).expand(-1, self.units)

        batch_size = x.shape[0]
        device = x.device

        # 1. Clip and normalize inputs
        x_clipped = torch.clamp(
            x, self.lb + 1e-9, self.ub - 1e-9
        )
        
        # 2. Calculate indices for piecewise segments
        # index = floor((x - lb + step) / step)
        indx = (
            (x_clipped - self.lb + self.step)
            / self.step
        ).long()
        indx = torch.clamp(indx, 0, self.num_buckets - 1)

        # 3. Construct activation vector (Step-based)
        range_vec = torch.arange(
            self.num_buckets, device=device
        ).view(1, 1, -1)
        # Expand index for broadcasting: [batch_size, units, 1]
        expand_indx = indx.unsqueeze(2)
        
        # Fully activated buckets (where range < index)
        # get the full step value
        activation_vector = torch.where(
            range_vec < expand_indx, 
            torch.tensor(self.step, device=device), 
            torch.tensor(0.0, device=device)
        )

        # 4. Handle the residue (delta) in the current bucket
        delta = (
            x_clipped - self.lb + self.step)
            - (indx.float() * self.step
        )
        
        # Use scatter to add the delta at the specific index
        # final_activation shape: [batch_size, units, num_buckets]
        final_activation = activation_vector.clone()
        final_activation.scatter_(2, expand_indx, delta.unsqueeze(2))

        # 5. Enforce non-negativity (Isotonic Constraint)
        # w = ReLU(v) ensures monotonic non-decreasing output
        weights = F.relu(self.w)
        if calibration_embedding is not None:
            weights = F.relu(self.w + calibration_embedding)

        # 6. Dot product fitting: 
        # y = sum(activation * w) + offset + bias
        logits = torch.sum(
            final_activation * weights, dim=2
        ) + self.residue + self.b
        
        return torch.sigmoid(logits)
\end{verbatim}
\end{small}
\end{document}